\documentclass[12pt]{iopart}

\usepackage[utf8]{inputenc}
\usepackage[T1,T2A]{fontenc}
\usepackage[ukrainian,russian,english]{babel}

\usepackage{iopams}
\usepackage{setstack}
\usepackage {graphicx}
\graphicspath{{fig/}}

\usepackage{cite}

\usepackage{etoolbox}

\patchcmd{\numparts}{\addtocounter{equation}{1}}{\refstepcounter{equation}}{}{}

\usepackage{comment}

\makeatletter
\def\@appendixstar{\@@par
 \ifnumbysec                        
 \@addtoreset{table}{section}        
 \@addtoreset{figure}{section}\fi
 \setcounter{section}{0}
 \setcounter{subsection}{0}
 \setcounter{subsubsection}{0}
 \setcounter{equation}{0}
 \setcounter{figure}{0}
 \setcounter{table}{0}
 \def\thesection{Appendix \Alph{section}}
 \def\theequation{\ifnumbysec
      \Alph{section}.\arabic{equation}\else
      \Alph{section}\arabic{equation}\fi}
 \def\thetable{\ifnumbysec
      \Alph{section}\arabic{table}\else
      A\arabic{table}\fi}
 \def\thefigure{\ifnumbysec
      \Alph{section}\arabic{figure}\else
      A\arabic{figure}\fi}
 \def\numparts{\refstepcounter{equation}%
     \setcounter{eqnval}{\value{equation}}%
     \setcounter{equation}{0}%
     \def\theequation{\ifnumbysec
     \Alph{section}.\arabic{eqnval}{\it\alph{equation}}%
     \else\arabic{eqnval}{\it\alph{equation}}\fi}}

\def\endnumparts{\def\theequation{\ifnumbysec
     \Alph{section}.\arabic{equation}\else
     \Alph{equation}\fi}%
     \setcounter{equation}{\value{eqnval}}}}
\makeatother

\usepackage{comment}

\usepackage{hyperref}
\hypersetup{
    colorlinks=true,        
    linkcolor=blue,          
    citecolor=blue,        
    filecolor=magenta,      
    urlcolor=blue           
}

\usepackage{color} 				
\definecolor{dgray}{rgb}{0.6,0.6,0.6}
\definecolor{dmag}{rgb}{0.6,0.0,0.6}
\usepackage[normalem]{ulem} 	

\definecolor{pink}{rgb}{1,0,0.9}

\newcommand{\eqref}[1]{{(\ref{#1})}}
\newcommand{\pdg}{{\phantom{\dagger}}}

\begin{document}
\title[]{\bf Magnetic phases and phase diagram of spin-1 condensate with quadrupole degrees of freedom}

\author{M S Bulakhov$^{1,2}$, A S Peletminskii$^{1,2}$, S V Peletminskii$^{1}$, and Yu~V~Slyusarenko$^{1,2}$}
\address{$^{1}$ Akhiezer Institute for Theoretical Physics, NSC KIPT, 61108 Kharkiv, Ukraine}
\address{$^{2}$ V.N.~Karazin Kharkiv National University, 61022 Kharkiv, Ukraine}
\ead{bulakh@kipt.kharkov.ua}
\ead{aspelet@kipt.kharkov.ua}



\begin{abstract}
We obtain and justify a many-body Hamiltonian of pairwise interacting spin-1 
atoms, which includes eight generators of the SU(3) group associated with spin 
and quadrupole degrees of freedom. It is shown that this Hamiltonian is valid for 
non-local interaction potential, whereas for local interaction specified by 
$s$-wave scattering length, the Hamiltonian should be bilinear in spin operators 
only (of the Heisenberg type). We apply the obtained Hamiltonian to study the 
ground-state properties and single-particle excitations of a weakly interacting 
gas of spin-1 atoms with Bose-Einstein condensate taking into account the 
quadrupole degrees of freedom. It is shown that the system under consideration 
can be in ferromagnetic, quadrupolar, and paramagnetic phases. The corresponding 
phase diagram is constructed and discussed. The main characteristics such as the 
density of the grand thermodynamic potential, condensate density, and 
single-particle excitation spectra modified by quadrupole degrees of freedom are 
determined in different phases.

\end{abstract}    
    
\maketitle

\section{Introduction}
Quantum spin systems provide an unique opportunity to probe novel states and corresponding phase transitions. Especially this concerns the high-spin magnets ($S>1/2$), which are specified by the non-Heisenberg Hamiltonian with competing interactions \cite{Nagaev1982} allowing to predict new quantum many-body states along with traditional ferromagnetic and antiferromagnetic orderings \cite{Matveev1974,Andreev1984,Papa1988}. 
The specific feature of high-spin systems is that they are described by additional multipole degrees of freedom (quadrupole in case of $S=1$, octupole for $S=3/2$, etc.) along with the dipolar spin vector. In recent years, there have been intensive studies of unconventional orderings in spin-1 \cite{Harada2002,Bernatska2009,Toth2012,Corboz2018,Kovalevsky2014} and spin-3/2 \cite{Kosmachev2011,Heinze2013, Kosmachev2015} magnets. Most of them are usually based on different lattice models.

Ultracold gases of interacting high-spin atoms represent another type of systems for the realization of various magnetic states.
For more than two decades, they attract much interest of scientific community due to very high control of the relevant physical parameters and possibility to simulate different effects and phenomena, which are difficult to probe in real materials. In particular, high-spin atoms loaded into optical lattices represent a powerful simulator of magnetic orderings previously studied theoretically for various lattices models \cite{Demler2003,Vekua2011,Lewenstein2011,Sotnikov2011,Sotnikov2015}. In this connection, it is natural to study the effects of quadrupole degrees of freedom and magnetic phases in interacting ultracold quantum gases of spin-1 atoms in a coherent state with a Bose-Einstein condensate (BEC). The earlier studies of the so-called spinor condensates \cite{HoPRL1998,JETP1998,Ohmi1998} do not take into account the multipole (quadrupole) degrees of freedom (see also the reviews \cite{UedaPhysRep,UedaRMP,Yuk2018}). This is mainly due to the specifics of describing the interaction in ultracold gases. Usually, in most studies, the interatomic interaction in these systems is specified by the $s$-wave scattering length describing the low-energy collisions of atoms \cite{Pethick2002,Pitaevskii2003}. In this case, for spin-1 atoms, it is sufficient to consider two interaction terms in the Hamiltonian with the corresponding coupling constants associated with scattering lengths. The first one is independent of spin operators, while the second term, bilinear in spin operators, describes the spin-spin interaction \cite{HoPRL1998}. The quadrupole degrees of freedom are not involved in such a description of the interaction effects in ultracold gases.

While the parameterization of interatomic interaction by the $s$-wave scattering length fairly well describes the relevant effects in ultracold gases, this approximation has some weak points. In particular, it does not take into account the finite range of interatomic potential that results in divergences of physical quantities such as the chemical potential or the ground state energy. To resolve the problem, the well-known renormalization procedure for the coupling constant is necessary \cite{Pethick2002,Pitaevskii2003}. However, even after applying this procedure, there remains some inconsistency in the description of a Bose gas with a condensate within the quadratic approximation based on the Bogoliubov model \cite{Bulakhov2018}. For spinor condensates, the scattering-length approximation does not allow taking into account the interaction effects associated with the quadrupole degrees of freedom and can lead to an incomplete structure of the single-particle excitation spectra \cite{PLA} so that the non-local character of interaction can not be ignored in some problems. The role of non-local interaction has been recently discussed in context of both ultracold Bose \cite{Bulakhov2018,Haas2018} and Fermi \cite{Hara2012,Caballero2013,Simonucci2011} gases . 

Therefore, if we refuse to describe the system of interacting spin-$S$ atoms by 
the scattering lengths, we should consider a more general interaction 
Hamiltonian, which includes the multipole operators (e.g., quadrupole ones in 
case of spin-1 atoms) along with spin operators. We show that for spin-1 systems, 
both spin and quadrupole operators can be treated through the prism of the SU(3) 
algebra on equal footing. In particular, three Gell-Mann generators specify three 
components of the spin operator, and the remaining five generators are associated 
with the quadrupole degrees of freedom. We obtain and justify a microscopic 
pairwise interaction Hamiltonian, which includes all eight generators of the 
SU(3) group. It is applied to study the ground-state structure and corresponding 
single-particle excitations of a weakly interacting gas of spin-1 atoms with BEC 
in an external magnetic field. We show how the quadrupole degrees of freedom 
(often called as hidden parameters) affect the main physical characteristics of 
the system under consideration. It is worth noting that quadrature squeezing 
of spin and nematic (quadrupole) variables can be realized in a spin-1 atomic 
BEC \cite{Hamley2012}. This makes it possible to study the components of 
the corresponding quadrupole matrix.  

The paper is organized as follows. In Sec.~2 we introduce the quadrupole degrees of freedom for quantum gases of spin-1 atoms by using the general definition of the single-particle density matrix and formalism of the SU(3) algebra. Next, we obtain and justify the many-body Hamiltonian of pairwise interaction for spin-1 atoms. In Sec.~4 we provide a derivation of the main equations for a weakly interacting Bose gas of spin-1 atoms with a Bose-Einstein condensate taking into account the quadrupole degrees of freedom. Sec.~4 deals with analysis of the ground-state structure and corresponding single-particle excitations of the system under consideration. In Sec.~5 we construct and discuss a phase diagram. Finally, we summarize our results in Sec.~6

\section{Quadrupole degrees of freedom in quantum gases} 

The necessity to introduce quadrupole degrees of freedom in spin-1 quantum gases can be seen by considering the single-particle density matrix $f_{\alpha\beta}({\bf p})={\rm Tr}\,\varrho a^{\dagger}_{{\bf p}\beta}a_{{\bf p}\alpha}$, where $\varrho$ is the equilibrium Gibbs statistical operator or non-equilibrium operator satisfying the Liouville equation and $a^{\dagger}_{{\bf p}\alpha}$, $a_{{\bf p}\alpha}$ are the creation and annihilation operators of bosonic spin-1 atoms, respectively. Being a $3\times 3$ matrix, the single-particle density matrix can be decomposed in the complete set of $3\times 3$ matrices,
\begin{eqnarray*}
f_{\alpha\beta}({\bf p})=f^{0}({\bf p})I_{\alpha\beta}+f^{a}({\bf p})\lambda_{\alpha\beta}^{a}, \quad a=1,\dots,8,
\end{eqnarray*}
where $I$ is the identity operator and $\lambda^{a}$ ($a=1,\dots,8$) are the Gell-Mann generators of the SU(3) Lie group (see Appendix A). The scalar $f^{0}$ and vector  $f^{a}$ decomposition coefficients are found to be
\begin{eqnarray*}
f^{0}=\frac13{\rm Tr}\,f({\bf p}), \quad f^{a}({\bf p})=\frac12{\rm Tr}\,f({\bf p})\lambda^a.
\end{eqnarray*}
Therefore, we see that eight parameters $f^{a}$ are necessary to describe a many-body system of spin-1 atoms. Moreover, these additional degrees of freedom are induced by a microscopic characteristic of an atom such as its spin.

The meaning of each Gell-Mann generator $\lambda^{a}$ can be understood by considering the realization of spin-1 operators in the Cartesian basis $\vert i \rangle$ ($i=x,\, y, \, z$) defined as $S^{i}\vert i \rangle=0 $, where $S^{i}$ are three components of the spin-1 operator. This basis satisfies the relations
\begin{equation}\label{eq:VectBasis}
\langle i\vert k\rangle=\delta_{ik}, \quad S^{i}\vert k\rangle =i\varepsilon_{ikl}\vert l\rangle, 
\end{equation}
so that the usual commutation relations for the components of the spin operator are valid, $[S{i},S^{k}]=i\varepsilon_{ikl}S^{l}$. From Eq.~(\ref{eq:VectBasis}), one obtains
\begin{eqnarray*}
\langle k\vert S^{i}\vert l\rangle\equiv (S^{i})_{kl}=-i\varepsilon_{ikl}, 
\end{eqnarray*}
that gives 
\begin{equation} \label{eq:SpinMatr}
S^{x}=\left(
  \begin{array}{ccc}
    0 & 0 & 0 \\ 
    0 & 0 & -i \\ 
    0 & i & 0 \\
  \end{array}
\right), \quad
S^{y}=\left(
  \begin{array}{ccc}
    0 & 0 & i \\
    0 & 0 & 0 \\
    -i & 0 & 0 \\
  \end{array}
\right), \quad
S^{z}=\left(
  \begin{array}{ccc}
    0 & -i & 0 \\ 
    i & 0 & 0 \\ 
    0 & 0 & 0 \\
  \end{array}
\right).
\end{equation}
Comparing these matrices with $\lambda^{a}$ (see Eq.~(\ref{eq:AGellMann})), we have $S^{x}\equiv\lambda^{7}$, $S^{y}=-\lambda^{5}$, and $S^{z}=\lambda^{2}$. Therefore, the subalgebra of three Gell-Mann generators coinciding with the components of the spin operator generates the SU(2) subgroup of the SU(3) group. To clarify the meaning of the remaining five matrices, we address the anticommutator relations given by Eq.~(\ref{eq:AAntiCom}), which yield 
\begin{eqnarray*}\label{eq:QuadOper}
\lambda^{1}=-\{S^{x},S^{y}\}, \quad \lambda^{3}=(S^{y})^{2}-(S^{x})^{2}, \quad \lambda^{4}=-\{S^{x},S^{z}\}, \nonumber \\
\lambda^{6}=-\{S^{y},S^{z}\}, \quad \lambda^{8}=\sqrt{3}(S^{z})^{2}-\frac2{\sqrt{3}}\,I.
\end{eqnarray*}
One can easily seen that the above five independent Gell-Mann generators form the traceless quadrupole matrix ${\cal Q}^{ik}\equiv S^{i}S^{k}+S^{k}S^{i}-(4/3)\,\delta_{ik}$ (see, e.g., Ref.~\cite{Corboz2018}), 
\begin{equation} 
\label{eq:QuadMatr}
{\cal Q}=\left(
  \begin{array}{ccc}
    -\lambda^{3}-\lambda^{8}/\sqrt{3} & -\lambda^{1} & -\lambda^{4} \\ 
    -\lambda^{1} & \lambda^{3}-\lambda^{8}/\sqrt{3} & -\lambda^{6} \\
    -\lambda^{4} & -\lambda^{6} & 2\lambda^{8}/\sqrt{3} \\
  \end{array}
\right)
\end{equation}
and we call them the quadrupole operators, which can be collected into a five-component vector $q^{b}=(-\lambda^{1},-\lambda^{3},-\lambda^{4},-\lambda^{6},\lambda^{8})$. Therefore, a many-body system of spin-1 atoms, in general case, is specified by the following spin and quadrupole operators:
\begin{equation}\label{eq:Spin}
S^{i}=\sum_{\bf p}a^{\dagger}_{{\bf p}\alpha}S^{i}_{\alpha\beta}a_{{\bf p}\beta}, \quad S^{i}=(S^{x}=\lambda^{7},S^{y}=-\lambda^{5},S^{z}=\lambda^{2})
\end{equation}
and 
\begin{equation}\label{eq:Quad}
Q^{b}=\sum_{\bf p}a^{\dagger}_{{\bf p}\alpha}q^{b}_{\alpha\beta}a_{{\bf p}\beta}, \quad q^{b}=(-\lambda^{1},-\lambda^{3},-\lambda^{4},-\lambda^{6},\lambda^{8}),
\end{equation}
which can be combined into a single eight-component operator, \begin{equation}\label{eq:SpinQuad}
\Lambda^{a}=\sum_{\bf p}a^{\dagger}_{{\bf p}\alpha}\lambda^{a}_{\alpha\beta}a_{{\bf p}\beta}, \quad a=1,\dots, 8.
\end{equation}

Therefore, taking into account Eqs.~(\ref{eq:Spin}) and (\ref{eq:Quad}) and employing the general rules for constructing binary quantities in the second quantization method, one can write a many-body Hamiltonian of pairwise interaction in the following form:  
\begin{eqnarray}
V_{UJK}=\frac1{2\mathcal{V}}\sum_{{\bf p}_{1},\ldots{\bf p}_{4}}U({\bf p}_{1}-{\bf p}_{3})a^{\dagger}_{{\bf p}_{1}\alpha}a^{\dagger}_{{\bf p}_{2}\beta}a_{{\bf p}_{3}\alpha}a_{{\bf p}_{4}\beta}\,\delta_{{\bf p}_{1}+{\bf p}_{2},\,{\bf p}_{3}+{\bf p}_{4}} \nonumber \\
+\frac1{2\mathcal{V}}\sum_{{\bf p}_{1},\ldots{\bf p}_{4}}J({\bf p}_{1}-{\bf p}_{3})a^{\dagger}_{{\bf p}_{1}\alpha}a^{\dagger}_{{\bf p}_{2}\beta}S^{i}_{\alpha \gamma}S^{i}_{\beta\delta}a_{{\bf p}_{3}\gamma}a_{{\bf p}_{4}\delta}\,\delta_{{\bf p}_{1}+{\bf p}_{2},\,{\bf p}_{3}+{\bf p}_{4}} \nonumber \\ 
+\frac1{2\mathcal{V}}\sum_{{\bf p}_{1},\ldots{\bf p}_{4}}K({\bf p}_{1}-{\bf p}_{3})a^{\dagger}_{{\bf p}_{1}\alpha}a^{\dagger}_{{\bf p}_{2}\beta}q^{b}_{\alpha \gamma}q^{b}_{\beta\delta}a_{{\bf p}_{3}\gamma}a_{{\bf p}_{4}\delta}\,\delta_{{\bf p}_{1}+{\bf p}_{2},\,{\bf p}_{3}+{\bf p}_{4}}, \label{eq:HamSU_2} 
\end{eqnarray}
where $\mathcal{V}$ is the volume of the system and $U({\bf p}_{1}-{\bf p}_{3})$, $J({\bf p}_{1}-{\bf p}_{3})$, and $K({\bf p}_{3}-{\bf p}_{1})$ are the Fourier transforms of the potential, spin-spin, and quadrupole-quadrupole  interaction energies, respectively. 

In order to justify the form of the above many-body Hamiltonian of pairwise interaction, which includes both spin and quadrupole degrees of freedom, we consider a collision of two spin-1 atoms. Let their interaction be specified by three coupling constants $g_{{\cal S}}$ (not yet scattering lengths; extension to the momentum-dependent Fourier transform of the real potential is trivial) corresponding to the total spin ${\cal S}=0,1,2$ channels. Then, the interaction Hamiltonian reads \cite{PLA}
\begin{equation}\label{eq:Inter}
V=c_{0}+c_{1}({\bf S}_{1}\cdot {\bf S}_{2})+c_{2}({\bf S}_{1}\cdot {\bf S}_{2})^{2},
\end{equation}
where 
\begin{eqnarray*}\label{eq:InterCoeff}
c_{0}=g_{1}+\frac13\left(g_{2}-g_{0}\right), \quad  c_{1}=\frac12\left(g_{2}-g_{1}\right), \quad c_{2}=\frac13\left(g_{0}-\frac{3g_{1}}2+\frac{g_{2}}2\right).
\end{eqnarray*}
Thus, in general case of three-channel scattering (corresponding to the total spin ${\cal S}=0,\,1,\,2$), the interaction Hamiltonian contains both bilinear and biquadratic terms in spin operators. The situation becomes completely different if the interatomic interaction is parameterized by the $s$-wave scattering lengths. In this case, we have only two coupling constants $g_{0}=4\pi h^{2}a_{0}/m$ and $g_{2}=4\pi h^{2}a_{2}/m$, where $a_{0}$ and $a_{2}$ are the scattering lengths corresponding to the total angular momentum ${\cal S}=0$ and ${\cal S}=2$, respectively. This is due to the fact that two identical spin-1 atoms in the $s$-state of relative motion (orbital momentum $l=0$) can not couple to form a state with the total spin ${\cal S}=1$ because this state is ruled out by the requirement for the wave function to be symmetric under the exchange of two atoms. In this case, the biquadratic term is expressed in terms of the bilinear term, $({\bf S}_{1}\cdot{\bf S}_{2})^{2}=2-({\bf S}_{1}\cdot{\bf S}_{2})$, so that the interaction Hamiltonian takes the form $V=\tilde{c}_{0}+\tilde{c}_{2}({\bf S}_{1}\cdot{\bf S}_{2}$), where $\tilde{c}_{0}=\frac13(g_{0}+2g_{2})$ and $\tilde{c}_{2}=\frac13(g_{2}-g_{0})$ \cite{PLA,HoPRL1998}. The comparison of Eqs.~(\ref{eq:Inter}), (\ref{eq:HamSU_2}) suggests that the quadrupole-quadrupole interaction should be expressed in terms of the biquadratic term in spin operators. Indeed, taking into account the following expressions for the squares of components of spin operators:
\begin{eqnarray*}
(S^{x})^{2}=\frac12\left(-\lambda^{3}-\frac1{\sqrt{3}}\lambda^{8}+\frac43\right), \nonumber \\
(S^{y})^{2}=\frac12\left(\lambda^{3}-\frac1{\sqrt{3}}\lambda^{8}+\frac43\right), \quad
(S^{z})^{2}=\frac13\left(\sqrt{3}\lambda^{8}+2\right), \label{eq:Spin-Gell}
\end{eqnarray*}
and for the mixed products of components:
\begin{eqnarray*}
S^{x}S^{y}=-\frac12\left(\lambda^{1}-i\lambda^{2}\right), \quad S^{y}S^{x}=-\frac12\left(\lambda^{1}+i\lambda^{2}\right), \nonumber \\
S^{x}S^{z}=-\frac12\left(\lambda^{4}-i\lambda^{5}\right), \quad S^{z}S^{x}=-\frac12\left(\lambda^{4}+i\lambda^{5}\right), \nonumber \\
S^{y}S^{z}=-\frac12\left(\lambda^{6}-i\lambda^{7}\right), \quad
S^{z}S^{y}=-\frac12\left(\lambda^{6}+i\lambda^{7}\right), \label{eq:Spin-Gell2} 
\end{eqnarray*}
one can show that the quantity $q^{b}_{\alpha\gamma}q^{b}_{\beta\delta}$ entering Eq.~(\ref{eq:HamSU_2}) can be written as 
\begin{equation}\label{eq:Biquadr}
\frac12 q^{b}_{\alpha\gamma}q^{b}_{\beta\delta}=S^{i}_{\alpha\sigma}S^{i}_{\beta\rho}S^{k}_{\sigma\gamma}S^{k}_{\rho\delta}+\frac12 S^{i}_{\alpha\gamma}S^{i}_{\beta\delta}-\frac43\delta_{\alpha\gamma}\delta_{\beta\delta}, 
\end{equation}
or, equivalently,
\begin{equation}\label{eq:Biquadr2} 
\frac12\lambda^{a}_{\alpha\gamma}\lambda^{a}_{\beta\delta}=S^{i}_{\alpha\gamma}S^{i}_{\beta\delta}+S^{i}_{\alpha\sigma}S^{i}_{\beta\rho}S^{k}_{\sigma\gamma}S^{k}_{\rho\delta}-\frac43\delta_{\alpha\gamma}\delta_{\beta\delta}. 
\end{equation}
Therefore, according to Eqs.~(\ref{eq:Biquadr}) and (\ref{eq:Biquadr2}), the appearance of quadrupole degrees of freedom in Eqs.~(\ref{eq:HamSU_2}) is equivalent to the fact that the interaction Hamiltonian contains the biquadratic term in spin operators. The Hamiltonian (\ref{eq:HamSU_2}) commutes with the total spin operator, see Eq.~(\ref{eq:Spin}), and, consequently, it has the SU(2) symmetry. If $J=K$, then $[V_{UJ},\Lambda^{a}]=0$ (see Eq.~(\ref{eq:SpinQuad})) and the interaction Hamiltonian becomes SU(3) symmetric,
\begin{eqnarray}
V_{UJ}=\frac1{2\mathcal{V}}\sum_{{\bf p}_{1},\ldots{\bf p}_{4}}U({\bf p}_{1}-{\bf p}_{3})a^{\dagger}_{{\bf p}_{1}\alpha}a^{\dagger}_{{\bf p}_{2}\beta}a_{{\bf p}_{3}\alpha}a_{{\bf p}_{4}\beta}\,\delta_{{\bf p}_{1}+{\bf p}_{2},\,{\bf p}_{3}+{\bf p}_{4}} \nonumber \\
+\frac1{2\mathcal{V}}\sum_{{\bf p}_{1},\ldots{\bf p}_{4}}J({\bf p}_{1}-{\bf p}_{3})a^{\dagger}_{{\bf p}_{1}\alpha}a^{\dagger}_{{\bf p}_{2}\beta}\lambda^{a}_{\alpha \gamma}\lambda^{a}_{\beta\delta}a_{{\bf p}_{3}\gamma}a_{{\bf p}_{4}\delta}\,\delta_{{\bf p}_{1}+{\bf p}_{2},\,{\bf p}_{3}+{\bf p}_{4}}. \label{eq:HamSU_3} 
\end{eqnarray}
This case can be realized by employing the Feshbach resonance. Quantum magnets 
with the enhanced SU(3) symmetry is of much current interest 
\cite{Bernatska2009,Bar'yakhtar2013}. Note that the formalism of the SU(3) Lie 
algebra was earlier employed to examine the spin-1 lattice models 
\cite{Bernatska2009,Bar'yakhtar2013,Onufrieva1981,Onufrieva1985} as well as 
to 
model the atomic excitations when studying the atomic squeezing for a system of 
three-level spin-1 atoms (${}^{87}{\rm Rb}$) in a cavity interacting with a 
radiation field \cite{Reboiro2013}.

In conclusion of this section, note that the contact (local) interatomic interaction specified by the $s$-wave scattering length does not allow taking into account the quadrupole degrees of freedom. In order to study their effect in ultracold gases, one should consider a more general interaction Hamiltonian with a non-trivial dependence of the Fourier transform of the interaction potential on momentum (the trivial dependence, i.e., $V({\bf p}=0)$ corresponds to the local interaction). In general, such interaction is not local and has a finite range. For indicated reasons, we present the Hamiltonian of pairwise interaction in the general form of Eqs.~(\ref{eq:HamSU_2}), not parameterizing it by the scattering lengths . Some comments on the weak points of local interaction are given at the end of Sec. 4.

\section{Bogoliubov model of a BEC with spin and quadrupole degrees of freedom}

In this section, we apply the SU(2) symmetric interaction Hamiltonian given by Eq.~(\ref{eq:HamSU_2}) to study the effect of quadrupole degrees of freedom on the ground state and corresponding excitations of a weakly interacting gas of spin-1 atoms with a Bose-Einstein condensate in a magnetic field. To this end, we extend the Bogolyubov model for a weakly interacting Bose gas \cite{Bogoliubov1947} to the following grand canonical Hamiltonian: 
\begin{equation}\label{eq:GrandHam1}
{\cal H}\equiv H-\mu N={\cal H}_{0}+V_{UJK},
\end{equation}
with  
\begin{equation}\label{eq:GrandHam2}
{\cal H}_{0}=\sum_{\bf p}a^{\dagger}_{{\bf p}\alpha}\left[(\varepsilon_{\bf p}-\mu)\delta_{\alpha\beta}- hS^{z}_{\alpha\beta}\right]a_{{\bf p}\beta}, \quad S^{z}_{\alpha\beta}\equiv\lambda^{2}_{\alpha\beta},
\end{equation}
where $\mu$ is the chemical potential, $N=\sum_{\bf p}a^{\dagger}_{{\bf p}\alpha}a^{\pdg}_{{\bf p}\alpha}$ is the particle number operator, $\varepsilon_{\bf p}={p^{2}/2m}$ is the kinetic energy of an atom, and
$h=g\mu_{B}B$ with $g$, $\mu_{B}$, and $B$ being the Land\'{e} hyperfine factor \cite{UedaPhysRep}, the Bohr magneton, and the external magnetic field directed along $z$-axis, respectively. The interaction Hamiltonian $V_{UJK}$ is given by Eq.~(\ref{eq:HamSU_2}). 

Since at zero temperature and weak interaction the number of bosons in a state with zero momentum is a macroscopic value, the Bogoliubov model proposes to treat the corresponding creation and annihilation operators as $c$-numbers neglecting normal and anomalous pair correlation functions \cite{Peletminskii2013,Stoof1994,Kondratenko1975,Peletminskii2010,Valatin1958,Girardeau1959,Evans1969,Coniglio1969}. The latter describe the pair-correlated bosons similar to Cooper pairs in BCS theory. Therefore, in all relevant operators of physical quantities, one should perform a replacement, $a_{0}^{\dagger}\to\sqrt{\cal V}\Psi^{*}_{\alpha}$ and $a_{0}\to\sqrt{\cal V}\Psi_{\alpha}$, where $\Psi_{\alpha}$ is the condensate wave function, which is a variational parameter. In particular, after this replacement in the Hamiltonian given by Eqs.~(\ref{eq:GrandHam1}), (\ref{eq:GrandHam2}), and (\ref{eq:HamSU_2}), we only keep the $c$-number terms and those which are quadratic in creation and annihilation operators with nonzero momentum but neglect the higher order terms containing three and four operators with finite momenta. The terms held in this way allow us to find the ground state and introduce the dispersion law for free quasiparticles (or single-particle excitations). The omitted terms describe the interaction between quasiparticles, which is not considered in this work. Therefore, the grand canonical Hamiltonian in the quadratic approximation reads 
\begin{eqnarray}
{\cal H}(\Psi)\approx {\cal H}^{(0)}(\Psi)+{\cal H}^{(2)}(\Psi), \label{eq:BogolHam}    
\end{eqnarray}
where ${\cal H}^{(0)}(\Psi)$ is its $c$-number part given by 
\begin{eqnarray*}\label{eq:cNumbHam}
\fl
\frac1{\mathcal{V}}{\cal H}^{(0)}(\Psi)=\frac{U(0)} 2(\Psi^{*}\Psi)^{2}+\frac{J(0)}2(\Psi^{*}S^{i}\Psi)^{2}+\frac{K(0)}2(\Psi^{*}q^{b}\Psi)^{2}-h(\Psi^{*}S^{z}\Psi)-\mu(\Psi^{*}\Psi)
\end{eqnarray*}
and ${\cal H}^{(2)}(\Psi)$ contains the terms quadratic in creation and annihilation operators with nonzero momentum,
\begin{eqnarray}
\fl
{\cal H}^{(2)}(\Psi)=\sum_{{\bf p}\neq0}a^{\dagger}_{{\bf p}\alpha}\left[(\varepsilon_{\bf p}-\mu)\delta_{\alpha\beta}- hS^{z}_{\alpha\beta}\right]a_{{\bf p}\beta} \nonumber \\ 
\fl
+
U(0)\sum_{{\bf p}\neq0}(\Psi^{*}\Psi)(a^{\dagger}_{\bf p}a_{\bf p})+\frac12\sum_{{\bf p}\neq 0}U({\bf p})\left[(a^{\dagger}_{\bf p}\Psi )(\Psi^{*}a_{\bf p})+(a^{\dagger}_{\bf p}\Psi)(a^{\dagger}_{-{\bf p}}\Psi )+{\rm h.c.}\right] \nonumber \\\fl
+J(0)\sum_{{\bf p}\neq 0}(\Psi^{*}S^{i}\Psi)(a^{\dagger}_{{\bf p}}S^{i}a_{\bf p}) 
	+\frac12
	\sum_{{\bf p}\neq 0}J({\bf p})\left[(a^{\dagger}_{\bf p}S^{i}\Psi)(\Psi^{*}S^{i}a_{\bf p})+(a^{\dagger}_{\bf p}S^{i}\Psi)(a^{\dagger}_{-{\bf p}}S^{i}\Psi)+{\rm h.c.}\right] \nonumber \\\fl
+ K(0)\sum_{{\bf p}\neq 0}
	(\Psi^{*}q^{b}\Psi)(a^{\dagger}_{{\bf p}}q^{b}a_{\bf p}) 
	+
	\frac12
	\sum_{{\bf p}\neq 0}K({\bf p})\left[(a^{\dagger}_{\bf p}q^{b}\Psi)(\Psi^{*}q^{b}a_{\bf p})+(a^{\dagger}_{\bf p}q^{b}\Psi)(a^{\dagger}_{-{\bf p}}q^{b}\Psi)+{\rm h.c.}\right],\nonumber\\ \label{eq:QuadrHam}
\end{eqnarray}
where we use the following notations, e.g., $(\Psi^{*}\Psi)\equiv\Psi^{*}_{\alpha}\Psi^{\phantom{*}}_{\alpha}$, $(a^{\dagger}_{\bf p}a^\pdg_{\bf p})\equiv a^{\dagger}_{{\bf p}\alpha}a^\pdg_{{\bf p}\alpha}$, $(\Psi^{*}S^{i}\Psi)\equiv(\Psi_{\alpha}^{*}S_{\alpha\beta}^{i}\Psi^{\phantom{*}}_{\beta})$. The corresponding Gibbs statistical operator in the quadratic approximation has the form
\begin{eqnarray*}\label{eq:GibbsOper}
w(\Psi)\approx\exp\left[\Omega(\Psi)-\beta{\cal H}(\Psi)\right].
\end{eqnarray*}
The grand thermodynamic potential $\Omega$ as a function of the reciprocal temperature, chemical potential, and condensate wave function is obtained from the normalization condition ${\rm Tr}\, w(\Psi)=1$ (trace is taken in space of occupation numbers of bosons with nonzero momentum), which yields
\begin{eqnarray*}\label{eq:GrandPot}
\Omega(\Psi)=\beta{\cal H}^{(0)}(\Psi)-\ln{\rm Tr}\left[ \exp(-\beta{\cal H}^{(2)}(\Psi))\right].
\end{eqnarray*}
According to the original Bogoliubov model \cite{Bogoliubov1947}, the variational parameter $\Psi_{\alpha}$ associated with the condensate wave function is found from the minimum condition for the $c$-number part of $\Omega$ assuming it to be the leading term, 
\begin{equation}\label{eq:GrandPotDens}
\fl
\varpi=\frac{\Omega}{\beta V}=\frac{U(0)}2(\Psi^{*}\Psi)^{2}+\frac{J(0)}2(\Psi^{*}S^{i}\Psi)^{2}+\frac{K(0)}2(\Psi^{*}q^{b}\Psi)^{2}-h(\Psi^{*}S^{z}\Psi)-\mu(\Psi^{*}\Psi).
\end{equation}
The introduced quantity $\varpi$, up to a sign, coincides with pressure $P=-\varpi$, which must be positive for thermodynamic stability (the thermodynamic potential is negative). The variation of Eq.~(\ref{eq:GrandPotDens}) with respect to $\Psi_{\alpha}^{*}$ with the subsequent setting to zero gives the following equation:
\begin{equation} \label{eq:MinCond}
\fl
\mu\zeta_{\alpha}-n_{0}U(0)\zeta_{\alpha}-n_{0}J(0)(\zeta^{*}S^{i}\zeta)S^{i}_{\alpha\beta} \zeta_{\beta}-n_{0}K(0)(\zeta^{*}q^{b}\zeta)q^{b}_{\alpha\beta}\zeta_{\beta}+hS^{z}_{\alpha\beta}\zeta_{\beta}=0,
\end{equation}
where we introduced the normalized spinor $\zeta_{\alpha}$, 
\begin{equation} \label{eq:NormSpinor}
\Psi_{\alpha}=\sqrt{n_{0}}\zeta_{\alpha}, \quad \zeta_{\alpha}^{\phantom{*}}\zeta_{\alpha}^{*}=1.
\end{equation}
Equation (\ref{eq:MinCond}), relating the condensate density $n_{0}$ to the chemical potential $\mu$, provides the minimum of the density of the grand thermodynamic potential. It allows for three types of solutions corresponding to ferromagnetic, quadrupolar, and paramagnetic phases. All these phases are studied in the next section.

\section{Ground-state structure and single-particle excitations}

To examine the ground state properties of the system under consideration, we address Eq.~(\ref{eq:MinCond}). These equations allow for three types of solutions (see Appendix B for details),
\begin{eqnarray}
{\rm F:}\qquad
&\bzeta=\frac{1}{\sqrt{2}}(1,i,0),
\quad
&\mu=n_{0}U(0)+n_{0}J(0)+\frac{1}{3}n_{0}K(0)-h, \label{eq:FChem} \\
{\rm Q:}\qquad
&\bzeta=(0,0,1)
,\quad 
&\mu=n_{0}U(0)+\frac{4}{3}n_{0}K(0), \label{eq:QChem} \\
{\rm P:}\qquad
&\bzeta=\frac{1}{2}(a,ib,0)
,\quad
&\mu=n_{0}U(0)+\frac{4}{3}n_{0}K(0), 
\label{eq:PChem} 
\end{eqnarray}
where
\begin{eqnarray}
a=\exp(i\varphi_{+})\sqrt{1+\frac{h}{c}}+\exp(i\varphi_{-})\sqrt{1-\frac{h}{c}}
\nonumber,\\
b=\exp(i\varphi_{+})\sqrt{1+\frac{h}{c}}-\exp(i\varphi_{-})\sqrt{1-\frac{h}{c}}. \label{eq:PCoeff}
\end{eqnarray}
In Eqs.~(\ref{eq:PCoeff}), $\varphi_{\pm}$ are real numbers and $c=n_{0}(J(0)-K(0))$. As we see below, all physical characteristics such as magnetization, pressure, single-particle excitation spectra  are independent of $\varphi_{\pm}$.

{\bf Ferromagnetic phase (F).} The ferromagnetic state is governed by Eqs.~(\ref{eq:FChem}). In this case the vector order parameter such as magnetization is given by
\begin{eqnarray}\label{eq:FMagn}
\langle S^{i}\rangle=(\Psi^{*}S^{i}\Psi)=n_{0}\delta_{iz}.
\end{eqnarray}
The quadrupole matrix (see Eq.~(\ref{eq:QuadMatr})) for the ferromagnetic phase reads
\begin{eqnarray*} \label{eq:FQuadr}
\langle {\cal Q}\rangle=(\Psi^{*}{\cal Q}\Psi)=n_{0}
\left(
  \begin{array}{ccc}
    -1/3 & 0 & 0 \\ 
    0 & -1/3 & 0 \\ 
    0 & 0 & 2/3 \\
  \end{array}
\right).
\end{eqnarray*}
The latter shows that the order parameter has rotational symmetry about the z axis since $\langle{\cal Q}^{xx}\rangle=\langle{\cal Q}^{yy}\rangle$. Next, from Eqs.~(\ref{eq:GrandPotDens}), (\ref{eq:NormSpinor}), (\ref{eq:FChem}) one obtains the density of thermodynamic potential, 
\begin{equation}\label{eq:FPot}
\varpi=-\frac{1}{2}\frac{(\mu+h)^{2}}{U(0)+J(0)+(1/3)K(0)}.
\end{equation}
For the ferromagnetic state to be stable, the density of thermodynamic potential must be negative (the pressure is positive). This yields the following necessary stability condition:
\begin{equation}\label{eq:FStab}
U(0)+J(0)+\frac{1}{3}K(0)>0.    
\end{equation}

In order to obtain the corresponding spectra of single-particle excitations, we return to the Hamiltonian quadratic in creation and annihilation operators given by Eq.~(\ref{eq:QuadrHam}). Employing the explicit form of spin and quadrupole operators and eliminating the chemical potential by using Eq.~(\ref{eq:FChem}), one finds
\begin{equation}\label{eq:FQuadrHam}
{\cal H}^{(2)}(n_{0})={\cal H}^{(2)}_{1}(n_{0})+{\cal H}^{(2)}_{2}(n_{0}),    
\end{equation}
where
\begin{equation}\label{eq:FHam1}
    {\cal H}^{(2)}_{1}(n_{0})
    =
    \sum_{{\bf p}\neq 0}
    \omega_{{\bf p} z}
    a^{\dagger}_{{\bf p}z}a_{{\bf p}z}
\end{equation}
and
\begin{equation}\label{eq:FHam2}
\fl
    {\cal H}^{(2)}_{2}(n_{0})=
    \sum_{{\bf p}\neq 0}a^{\dagger}_{{\bf p}\alpha}
    A_{\alpha\beta}
    a_{{\bf p}\beta}
+
    \frac12
    \sum_{{\bf p}\neq 0}
    a^{\dagger}_{{\bf p}\alpha}B_{\alpha\beta}
    a^{\dagger}_{-{\bf p}\beta}
+
    \frac12
    \sum_{{\bf p}\neq 0}
    a_{{\bf p}\alpha}B^{*}_{\alpha\beta}a_{-{\bf p}\beta}, \quad \alpha,\,\beta=x,\,y.
\end{equation}
Here $A_{\alpha\beta}$ and $B_{\alpha\beta}$ are Hermitian ($A=A^{\dagger}$) and symmetric ($B=B^{T}$) $2\times2$ matrices, respectively, with the following matrix elements:
\begin{eqnarray*}
A_{xx}=A_{yy}
=\varepsilon_{\bf p}+h+\frac{1}{2} n_0 U({\bf p})-n_0 J(0)+\frac{1}{2}n_0 J({\bf p})+\frac{7}{6}n_0 K({\bf p}), \nonumber \\
A_{xy}=A_{yx}^{*}=i\left(h-\frac{1}{2}n_0 U({\bf p})-n_0 J(0)-\frac{1}{2}n_0 J({\bf p})+\frac{5}{6}n_0 K({\bf p})\right), \label{eq:FMatrix1}
\end{eqnarray*}
and
\begin{eqnarray*} \label{eq:FMatrix2}
    B_{xx}=-B_{yy}=\frac{1}{2} n_0 U({\bf p})-\frac{1}{2} n_0 J({\bf p}) + \frac{1}{6} n_0 K({\bf p}), \quad B_{xy}=B_{yx}=iB_{xx}.
\end{eqnarray*}
In general, the Hamiltonian quadratic in creation and annihilation operators should be diagonalized as the whole operator. However, according to Eqs.~(\ref{eq:FHam1}), (\ref{eq:FHam2}), the creation and annihilation operators in both parts of the total Hamiltonian do not mix with each other, so that ${\cal H}_{1}^{(2)}(n_{0})$ and ${\cal H}_{2}^{(2)}(n_{0})$ can be diagonalized separately. However, ${\cal H}_{1}^{(2)}(n_{0})$ has already a diagonal form with the following dispersion law:
\begin{equation}\label{eq:FSpectrz}
    \omega_{{\bf p}z}=\varepsilon_{\bf p}+h+n_{0}\left[J({\bf p})
	-J(0)\right]+n_{0}\left[K({\bf p})-K(0)\right].
   \end{equation}
Therefore, we only need to diagonalize ${\cal H}_{2}^{(2)}(n_{0})$. This can be done by employing the general Bogoliubov procedure \cite{Bogol_Intr} for diagonalizing a quadratic form given by Eq.~(\ref{eq:FHam2}). Following it, ${\cal H}_{2}^{(2)}(n_{0})$ can be reduced to the diagonal form,
\begin{eqnarray*}
U{\cal H}_{2}^{(2)}(n_{0})U^\dagger=\sum_{{\bf p}\neq 0}\sum_{\gamma=x,y}\omega_{{\bf p}\gamma}a^{\dagger}_{{\bf p}\gamma}a_{{\bf p}\gamma}+{\cal E}_{0},
\end{eqnarray*}
where $U$ is the canonical unitary transformation \cite{AkhPel,PLA} and ${\cal 
E}_{0}$ determines the ground state thermodynamic potential including the 
contribution from the quadratic terms in creation and annihilation operators. 
This quantity is not considered in the present study. The single-particle 
excitation energies $\omega_{{\bf p}x}$ and $\omega_{{\bf p}y}$ are given by
\begin{equation}\label{eq:FSpectrx}
\omega_{{\bf p}x}=\varepsilon_{\bf p}+2h+2n_{0}\left[K({\bf p})-J(0)\right], 
\end{equation}
and
\begin{equation}\label{eq:FSpectry}
\omega_{{\bf p}y}=\left[\varepsilon_{\bf p}^{2}+2\varepsilon_{\bf p}n_{0}\left(U({\bf p})+J({\bf p })+\frac{1}{3}K({\bf p })\right)\right]^{1/2}.
\end{equation}
Therefore, the ferromagnetic state is characterized by three branches of the single-particle excitation spectrum. The first two branches, given by Eqs.~(\ref{eq:FSpectrz}) and (\ref{eq:FSpectrx}), do not depend on the interaction amplitude $U({\bf p})$ and describe the spin-quadrupole waves. Both of them have the activation energy, or the energy gap. The third branch of the spectrum, determined by Eq.~(\ref{eq:FSpectry}), represents the gapless Bogoliubov mode modified by spin-spin and quadrupole-quadrupole interactions. At small momenta, it becomes linear in ${\bf p}$, $\omega_{{\bf p}y}\approx sp$, where 
\begin{equation}
s=\sqrt{\frac{n_{0}}{m}\left(U(0)+J(0)+\frac{1}{3}K(0)\right)}.  
\end{equation}
is the speed of sound, which coincides with its general definition $s=\sqrt{\partial P/\partial\rho_{0}}$, where $\rho_{0}=mn_{0}$ is the mass density and $P=-\varpi$ is the pressure expressed in terms of $\rho_{0}$ according to Eqs.~(\ref{eq:FChem}) and (\ref{eq:FPot}). The requirement for the speed of sound to be a real number results in the stability condition given by Eq.~(\ref{eq:FStab}).

{\bf Quadrupolar phase (Q).} The quadrupolar phase is characterized by a zero magnetization $\langle S^{i} \rangle = 0$, but it breaks the spin-rotation symmetry by developing an anisotropy in spin fluctuations \cite{Corboz2018}. Indeed, the state vector $\bzeta$, given by Eq.~(\ref{eq:QChem}), simultaneously suppresses the magnetization,
\begin{eqnarray}\label{eq:QMagn}
\langle S^{i}\rangle=(\Psi^{*}S^{i}\Psi)=0
\end{eqnarray}
and breaks the spin-rotation symmetry,
\begin{eqnarray*}\label{eq:QFluct}
\langle(S^{z})^{2}\rangle=0, \quad \langle(S^{x})^{2}\rangle=\langle(S^{y})^{2}\rangle=n_0.
\end{eqnarray*}
In this case, for the quadrupole matrix, we have
\begin{eqnarray*} \label{eq:QQuadr}
\langle {\cal Q}\rangle=(\Psi^{*}{\cal Q}\Psi)=n_{0}\left(\frac{2}{3}\delta_{ik}-2e_{i}e_{k}\right),
\end{eqnarray*}
where a unit vector $e_{z}=\pm 1$ perpendicular
to the plane of fluctuations is called a director. This indicates that the spin vector fluctuates in the $xy$ plane. According to Eq.~(\ref{eq:GrandPotDens}), (\ref{eq:NormSpinor}), and (\ref{eq:QChem}), the density of thermodynamic potential is
\begin{equation} \label{eq:QPot}
\varpi=-\frac{1}{2}\frac{\mu^{2}}{U(0)+(4/3)K(0)}.
\end{equation}
Note that in the quadrupolar phase, the density of thermodynamic potential is independent of the external magnetic field $h$. Since $\varpi$ should be negative, the stability condition reads 
\begin{equation}\label{eq:QStab}
U(0)+(4/3)K(0)>0.    
\end{equation}

As before, the corresponding spectra of single-particle excitations are obtained from the general quadratic Hamiltonian given by Eq.~(\ref{eq:QuadrHam}). Using Eqs.~(\ref{eq:QChem}) for the quadrupolar phase, the Hamiltonian can be written as a sum of two terms 
(see Eq.~(\ref{eq:FQuadrHam})), where
\begin{equation}\label{eq:QQudrHam1}
{\cal H}_{1}^{(2)}(n_{0})=\sum_{{\bf p}\neq 0}\alpha_{{\bf p}z}a^{\dagger}_{{\bf p}z}a_{{\bf p}z}+\frac{1}{2}\sum_{{\bf p}\neq 0}\beta_{{\bf p}z}\left[a^{\dagger}_{{\bf p}z}a^{\dagger}_{-{\bf p}z}+a_{{\bf p}z}a_{-{\bf p}z}\right] 
\end{equation}
with
\begin{eqnarray*}\label{eq:QCoeff} 
\alpha_{{\bf p}z}=\varepsilon_{{\bf p}}+\beta_{{\bf p}z}, \quad \beta_{{\bf p}z}=n_{0}\left(U({\bf p})+\frac{4}{3}K({\bf p})\right).
\end{eqnarray*}
Once the Hamiltonian has the form of Eq. (\ref{eq:QQudrHam1}), the quasiparticle energy can be written immediately as $\omega_{{\bf p}z}=(\alpha_{{\bf p}z}^{2}-\beta_{{\bf p}z}^{2})^{1/2}$ \cite{AkhPel} or 
\begin{equation}\label{eq:QSpectrz}
\omega_{{\bf p}z}=
\left[\varepsilon_{\bf p}^{2}+2\varepsilon_{\bf p}n_{0}\left(U({\bf p})+\frac{4}{3}K({\bf p })\right)\right]^{1/2}.
\end{equation}
The second part of the Hamiltonian ${\cal H}^{(2)}_{2}(n_{0})$, which can again be diagonalized separately, has the form of Eq.~(\ref{eq:FHam2}) in which the $2\times 2$ matrices $A=A^{\dagger}$ and $B=B^{T}$ have the following matrix elements:
\begin{eqnarray*}\label{eq:QMatrixA}
A_{xx}=A_{yy}=\varepsilon_{{\bf p}}+n_{0}J({\bf p})-2n_{0}   K(0)+n_{0}K({\bf p}), \quad
A_{xy}=A_{yx}^{*}=ih
\end{eqnarray*}
and
\begin{eqnarray*}\label{eq:QMatrixB}
B_{xx}=B_{yy}=n_{0}K({\bf p})-n_{0}J({\bf p}), \quad
B_{xy}=B_{yx}=0.
\end{eqnarray*}
The general diagonalization procedure \cite{Bogol_Intr} applied to ${\cal H}^{(2)}_{2}(n_{0})$ reduces the corresponding operator to the Hamiltonian of free quasiparticles with the following dispersion laws:
\begin{equation}\label{eq:QSpectrx_y}
\fl
\omega_{{\bf p}x,y}=\left[\left(\varepsilon_{\bf p}+n_{0} J({\bf p })+n_{0}K({\bf p })-2n_{0}K(0)\right)^2-\left(n_{0} J({\bf p })-n_{0}K({\bf p })\right)^2\right]^{1/2}\pm h.
\end{equation}
Therefore, the quadrupolar phase under consideration is specified by three types of single-particle excitations. The first type, given by Eq.~(\ref{eq:QSpectrz}), is the Bogoliubov gapless mode modified by quadrupole degrees of freedom. It is independent of applied magnetic field $h$ and describes the phonon (density) excitations at small momentum, $\omega_{{\bf p}z}\approx sp$, where the speed of sound is given by
\begin{equation} \label{eq:QSpeed}
s=\sqrt{\frac{n_{0}}{m}\left(U(0)+\frac{4}{3}K(0)\right)}.
\end{equation}
In contrast to the ferromagnetic phase, the interaction amplitude $J({\bf p})$ of spin-spin interaction does not contribute to this mode. Note that, just as in the ferromagnetic case, a speed of sound determined by Eq.~(\ref{eq:QSpeed}) is consistent with its general definition, $s=\sqrt{\partial P/\partial\rho_{0}}$. For the system to be stable, the phonon mode (speed of sound) must be real. This implies the stability condition in the form of Eq.~(\ref{eq:QStab}). The other two modes (see Eq.~(\ref{eq:QSpectrx_y})), differing only by the sign of the magnetic field, do not contain the interaction amplitude $U({\bf p})$ and describe the spin-quadrupole waves. In the general case, they have a gap $h$ that vanishes in the absence of a magnetic field.

{\bf Paramagnetic phase (P).} Paramagnetic phase does not support any magnetization in the absence of an external magnetic field. Indeed, Eqs.~(\ref{eq:NormSpinor}) and (\ref{eq:PChem}) yield 
\begin{eqnarray}\label{eq:PMagn}
\langle S^{i}\rangle=(\Psi^{*}S^{i}\Psi)=\frac{h}{J(0)-K(0)} \delta_{iz}.
\end{eqnarray}
For the paramagnetic state, the quadrupole tensor describing the anisotropy of spin fluctuations, according to Eq.~(\ref{eq:QuadMatr}), reads
\begin{equation}
 \label{eq:QQTensor}
\fl
\langle {\cal Q}\rangle=(\Psi^{*}{\cal Q}\Psi)=
n_{0} 
\left(
  \begin{array}{ccc}
    -1/3-\gamma\cos(\phi_{+}-\phi_{-}) & \gamma\sin(\phi_{+}-\phi_{-}) & 0 \\ 
    \gamma\sin(\phi_{+}-\phi_{-}) & -1/3+\gamma\cos(\phi_{+}-\phi_{-}) & 0 \\ 
    0 & 0 & 2/3 \\
  \end{array}
\right),
\end{equation}
where $\gamma=\sqrt{1-(h/c)^{2}}$. Since $\phi_{\pm}$ are arbitrary real numbers, one can choose them to be zero without loss of generality. In this case, the off-diagonal matrix elements in Eq.~(\ref{eq:QQTensor}) vanish and the spin fluctuations are anisotropic in the $xy$ plane because $\langle {\cal Q}^{xx}\rangle\neq \langle {\cal Q}^{yy}\rangle$. As we see below, all physical quantities such as magnetization, pressure, and quasiparticle energy are independent of $\phi_{\pm}$. In particular, the density of thermodynamic potential is found from Eqs.~(\ref{eq:GrandPotDens}), (\ref{eq:NormSpinor}), and (\ref{eq:PChem}),
\begin{equation}\label{eq:PPot}
\varpi=-\frac{1}{2}\left(\frac{h^2}{J(0)-K(0)}+    \frac{\mu^{2}}{U(0)+(4/3)K(0)}
    \right).
\end{equation}
For the paramagnetic phase to be stable, at least one of the following inequalities must hold:
\begin{equation}\label{eq:PStab}
    J(0)>K(0)
    \quad
    \quad
    U(0)+(4/3)K(0)>0.
\end{equation}
Moreover, in accordance with Eq.~(\ref{eq:PCoeff}), one more condition 
\begin{equation}\label{PFieldRestr}
h\leq n_0|J(0)-K(0)|
\end{equation}
imposing a restriction on the magnetic field is necessary for the paramagnetic phase to be realized.

As in the previous two cases, the total quadratic Hamiltonian corresponding to paramagnetic phase splits into two operators ${\cal H}_{1}^{(2)}(n_{0})$ and ${\cal H}_{2}^{(2)}(n_{0})$, which can be diagonalized independently one of another. The first one, ${\cal H}_{1}^{(2)}(n_{0})$, has the form of Eq.~(\ref{eq:QQudrHam1}) but with a different definition of $\alpha_{{\bf p}z}$ and $\beta_{{\bf p}z}$:
\begin{eqnarray*}
\alpha_{{\bf p}z}=\varepsilon_{\bf p}+n_{0}(J({\bf p})+K({\bf p})-2K(0)), \quad
\beta_{{\bf p}z}=n_{0}\gamma\left(K({\bf p})-J({\bf p})\right).
\end{eqnarray*}
The corresponding single-particle excitation energy of the diagonalized Hamiltonian is $\omega_{{\bf p}z}=(\alpha^{2}_{{\bf p}z}-\beta^{2}_{{\bf p}z})^{1/2}$, or explicitly:
\begin{equation}\label{eq:PSpectrz}
\fl
\omega_{{\bf p}z}=\left[\left(\varepsilon_{\bf p}+n_{0}J({\bf p })+n_{0}K({\bf p })-2n_{0}K(0)\right)^{2}-\gamma^{2}\left(n_{0}J({\bf p })-n_{0}K({\bf p })\right)^{2}\right]^{1/2}. \\
\end{equation}
The structure of the second operator ${\cal H}_{2}^{(2)}(n_{0})$ is given by Eq.~(\ref{eq:FHam2}) in which the matrix elements of Hermitian, $A=A^{\dagger}$, and symmetric, $B=B^{T}$, matrices are the following:  
\begin{eqnarray*}
A_{\rho\rho}=\varepsilon_{\bf p}+\frac{n_0}{2}\left(U({\bf p})+J({\bf p})+\frac{7}{3}K({\bf p})-2K(0)\right)\\\pm
        \frac{n_{0}\gamma}{2}\left(U({\bf p})-J({\bf p})+\frac{1}{3}K({\bf p})+2K(0)\right), \\
        A_{xy}=A_{yx}^{*}=-\frac{ihn_0}{2c}\left(U({\bf p})+J({\bf p})-\frac{5}{3}K({\bf p})+2K(0)\right),
\end{eqnarray*}
and
\begin{eqnarray*}
B_{\rho\rho}=\pm\frac{n_0}{2}\left(U({\bf p})+J({\bf p})+\frac{1}{3}K({\bf p})\right)+\frac{n_{0}\gamma}{2}\left(U({\bf p})-J({\bf p})
        +\frac{7}{3}K({\bf p})\right), \\
        B_{xy}=B_{yx}=\frac{ihn_0}{2c}\left(U({\bf p})+J({\bf p})+\frac{1}{3}K({\bf p})\right),
\end{eqnarray*}
where a plus sign in $A_{\rho\rho}$ and $B_{\rho\rho}$ ($\rho=x,y$) corresponds to $A_{xx}$ and $B_{xx}$, whereas a minus sign to $A_{yy}$ and $B_{yy}$. Next, following again the general procedure for diagonalizing quadratic Hamiltonians \cite{Bogol_Intr}, we arrive at two additional modes of  single-particle excitations,
\begin{eqnarray*}
\omega_{{\bf p}x,y}=\left(\varepsilon_{\bf p}^{2}+D\varepsilon_{\bf p}+F\pm\sqrt{G\varepsilon_{\bf p}^2+L\varepsilon_{\bf p}+F^2} \right)^{1/2}
\end{eqnarray*}
where
\begin{eqnarray*}
D=n_{0}\left(U({\bf p})+J({\bf p})+\frac{7}{3}K({\bf p})-2K(0)\right), \\
F=2n_{0}^{2}\left(K({\bf p})-K(0)\right)^2\left(\gamma^{2}\frac{J({\bf p})-K({\bf p})}{K({\bf p})-K(0)}
+1\right), \\\fl
G=n_{0}^{2}\left(U({\bf p})-J({\bf p})+\frac{1}{3}K({\bf p})+2K(0)\right)^2\\+4n_{0}^{2}\frac{h^2}{c^2}\left(J({\bf p})-K({\bf p})\right)\left(U({\bf p})-\frac{2}{3}K({\bf p})+2K(0)\right), \\\fl
L=-4n_{0}^{3}\left(K({\bf p})-K(0)\right)^{2}\left[U({\bf p})+J({\bf p})-\frac{5}{3}K({\bf p})+2K(0)
\right.
\\
  \left.
 +\gamma^{2}\frac{J({\bf p})-K({\bf p})}{K({\bf p})-K(0)}\left(U({\bf p})-J({\bf p})-\frac{5}{3}K({\bf p})+4K(0)\right)\right].
\end{eqnarray*}

Note that the first mode, given by Eq.~(\ref{eq:PSpectrz}), is always gapful with a gap vanishing at zero magnetic field ($\gamma=1$). It is independent of the interaction amplitude
$U({\bf p})$ and describes the spin-quadrupole oscillations. As for the other two branches $\omega_{{\bf p}x,y}$, despite the fact that they are linear in momentum (in the limit $p\to 0$), they do not provide a speed of sound consistent with its general definition $s=\sqrt{\partial P/\partial\rho_{0}}$, where $P=-\varpi$ (see Eqs.~(\ref{eq:PPot}) and (\ref{eq:PChem})). Therefore, the density and spin-quadrupole excitations are coupled in these modes. It is worth noting that at zero magnetic field, all physical characteristics of paramagnetic phase such as the density of thermodynamic potential, chemical potential, single-particle excitation spectra coincide with those for the quadrupolar phase. 

In the case when the interatomic interaction is SU(3) symmetric, i.e., $J({\bf p})=K({\bf p})$, all the obtained results agree with the earlier study \cite{PLA}. Moreover, the paramagnetic phase in this case can not occur as the ground state due to the difference $J(0)-K(0)$ in the denominators of the relevant physical quantities (see Eqs.~(\ref{eq:PChem}) and (\ref{eq:PPot})). In case of SU(2) symmetric Hamiltonian, which is bilinear in spin operators and does not include the quadrupole degrees of freedom ($K({\bf p})=0$), the obtained results reproduce the studies with non-local interaction \cite{JETP1998,PhysicaA}. If the interaction is taken to be of the contact type (local interaction), 
\begin{equation} \label{eq:LocalPot}
\fl
U({\bf p})=U(0)=\frac{g_{0}+2g_{2}}{3}, \quad J({\bf p})=J(0)=\frac{g_{2}-g_{0}}{3}, \quad g_{0,2}=\frac{4\pi\hbar^{2}}{m}a_{0,2},
\end{equation}
where $a_{0,2}$ are the $s$-wave scattering lengths corresponding to the total spin 0 or 2 of two colliding spin-1 atoms, the results are also in agreement with those obtained in Refs.~\cite{Ohmi1998,HoPRL1998,UedaPhysRep}. 

Some comments should be made regarding the local interaction in theory of a weakly interacting Bose gas with BEC. Although the scattering-length approximation has been proved to be a powerful tool to describe the interaction effects in ultracold gases, it has some weak points. In particular, it does not take into account the finite range of the interaction potential and, as a consequence, the сorresponding integrals governing the ground state energy or the chemical potential diverge at zero momentum, so that the well-known artificial renormalization procedure \cite{Pethick2002,Pitaevskii2003} is required to remove the divergences occurring in the terms quadratic in the creation and annihilation operators (see Ref.~\cite{Peletminskii2017} for higher order terms). Moreover, even after this procedure, the general equation in the consistent quadratic approximation (based on the Bogoliubov model) providing the minimum of the grand thermodynamic potential has no solution. At the same time, this equation has a solution for the potentials of a finite range \cite{Bulakhov2018}. Besides that, as it was shown in Sec.~1 (see also Ref.~\cite{PLA}), the scattering-length approximation does not allow one to take into account the possible manifestation of the quadrupole degrees of freedom in the effects of interatomic interaction in ultracold gases. Finally, even if the quadrupole degrees of freedom are not taken into account in the interaction Hamiltonian, $K({\bf p})=K(0)=0$, the single-particle excitation spectra have an incomplete structure. Indeed, Eq.~(\ref{eq:FSpectrz}) shows that under conditions given by Eqs.~(\ref{eq:LocalPot}), the corresponding single-particle excitation energy becomes fully independent of the interaction parameters, see also  \cite{Ohmi1998,HoPRL1998,UedaPhysRep}. Meanwhile, it is clear that when describing a system of interacting atoms, the interaction parameters should determine the dispersion law of quasiparticles. A similar situation arises in the SU(3) symmetric case, when $J({\bf p})=K({\bf p})$ \cite{PLA}. The role of non-local interaction in physics of ultracold gases has been recently examined in a number of studies \cite{Bulakhov2018,Haas2018,Hara2012,Caballero2013,Simonucci2011}.

\section{Stability and phase diagram}

In this section we illustrate the emerging magnetic phases for a weakly interacting gas of spin-1 atoms with Bose-Einstein condensate by plotting corresponding diagrams in the plane of dimensionless interaction parameters. This can be done by considering the necessary stability conditions (see Eqs.~(\ref{eq:FStab}), (\ref{eq:QStab}), and (\ref{eq:PStab})) and comparing the densities of the thermodynamic potential (see Eqs.~(\ref{eq:FPot}), (\ref{eq:QPot}), and (\ref{eq:QPot})) for each many-body state. 

\begin{figure}[htbp]
\centering
\includegraphics[width=85mm]{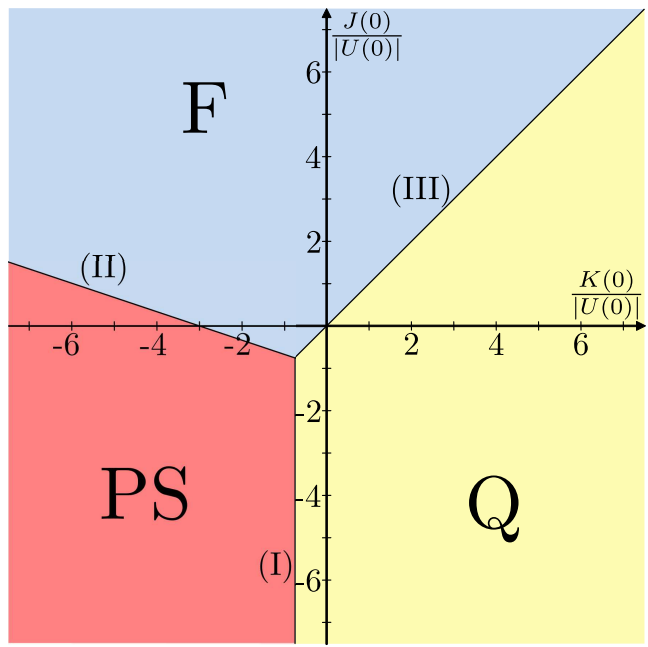}
\caption{Phase diagram with ferromagnetic (F), quadrupolar (Q) phases, and forbidden (phase separation) (PS) region with no stable condensate solution. 
The transition lines (I) $J(0)=-U(0)-(1/3)K(0)$ and (II) $K(0)=-(3/4)U(0)$ correspond to the boundaries of the stability conditions (see Eqs.~\eqref{eq:FStab} and \eqref{eq:QStab}). On the line (III) $J(0)=K(0)$, the densities of the thermodynamic potential (see Eqs.~(\ref{eq:FPot}) and (\ref{eq:QPot})) are equal.
}
\label{fig:ph_diagram_1}
\end{figure}

Fig.~\ref{fig:ph_diagram_1} shows the phase diagram when the density of the thermodynamic potential is considered as a function of the condensate density $\varpi=\varpi(n_{0})$ (the chemical potential is eliminated for each phase by employing Eqs.~(\ref{eq:FPot})-(\ref{eq:PPot})) The condensate density is assumed to be fixed. In this case, the paramagnetic phase loses to both ferromagnetic (F) and quadrupolar (Q) states. In the red region, the condensate is unstable since the thermodynamic potential is positive (the pressure is negative). Conventionally, we call it as the phase separation (or forbidden) region (PS). This diagram demonstrates two interesting features of the system under consideration. The SU(3) symmetric case is realized on the line $J(0)=K(0)$, where the ferromagnetic and quadrupole phases are equally favorable. When the quadrupole degrees of freedom are not involved in the interaction Hamiltonian, $K(0)=0$, the preference of a particular phase is determined only by the sign of the spin-spin interaction $J(0)$, like in the usual Heisenberg model.

\begin{figure}[htbp]
	\centering
	\includegraphics[width=\linewidth]{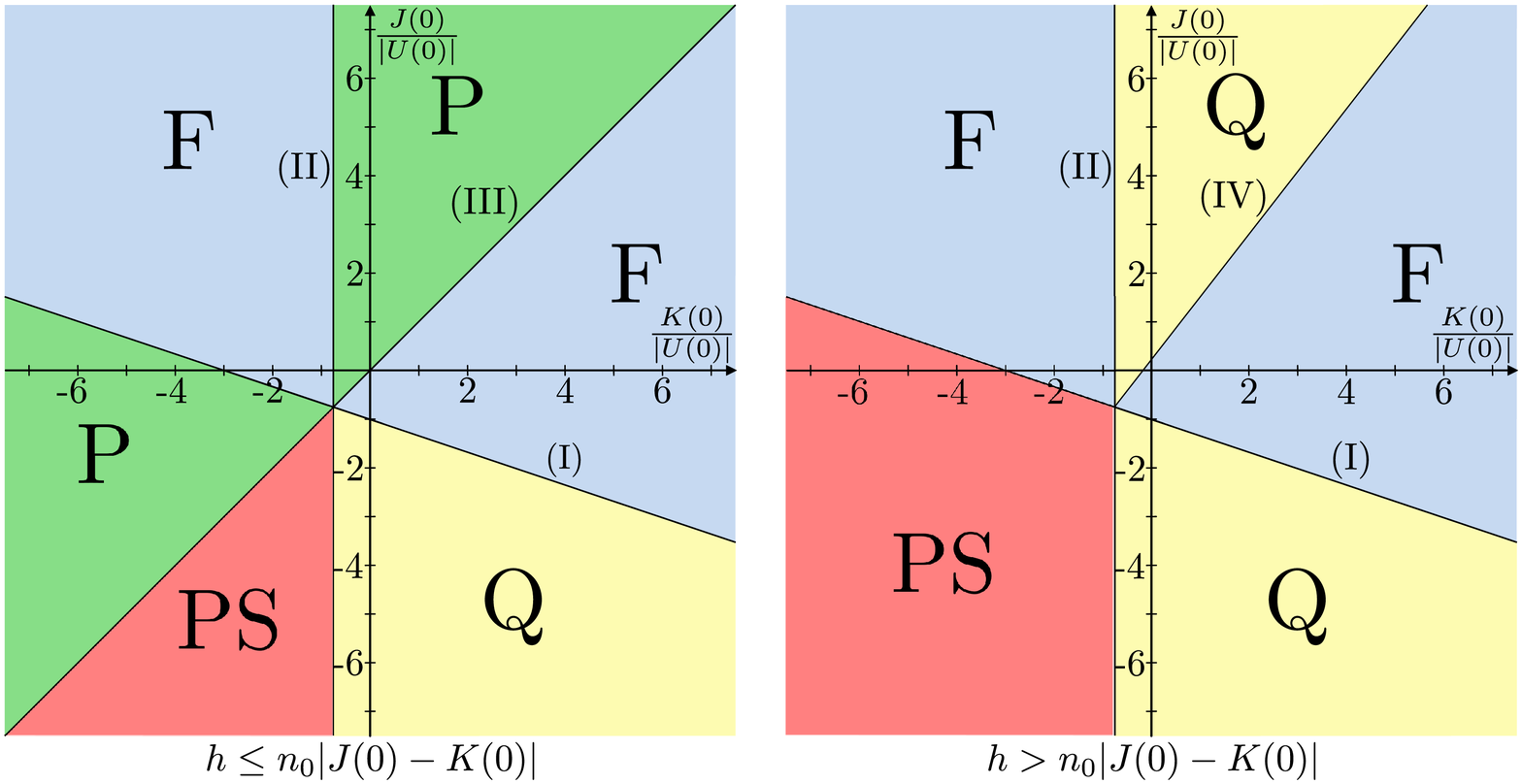}
	\caption{Phase diagrams with ferromagnetic (F), quadrupolar (Q), paramagnetic (P) phases, and forbidden (phase separation) (PS) region  with no stable condensate solution. 
	The transition lines (I) $J(0)=-U(0)-(1/3)K(0)$ and (II) $K(0)=-(3/4)U(0)$, and (III) $J(0)=K(0)$ correspond to the boundaries of the stability conditions (see Eqs.~\eqref{eq:FStab}, \eqref{eq:QStab}, and \eqref{eq:PStab}, respectively). 
	On the line (IV) $J(0)=\left[1+\frac43\frac{2\mu h+h^2}{\mu^2}\right]K(0)+\frac{2\mu h+h^2}{\mu^2} U(0)$, the densities of the thermodynamic potential are equal for (F) and (Q) phases (see Eqs.~(\ref{eq:FPot}) and (\ref{eq:QPot}), respectively).  
	}
	\label{fig:ph_diagram_2}
\end{figure}

If the system is described in terms of chemical potential, $\varpi=\varpi(\mu)$, there exist regions of parameters in which the paramagnetic phase is most favourable (see Fig.~\ref{fig:ph_diagram_2}, the chemical potential $\mu$ is fixed and assumed to be positive). Since the possibility of this phase to exist is determined by the magnitude of the magnetic field ($h<n_0 | J (0) -K (0) |$), there must be two diagrams. The left diagram of Fig.~2 corresponds to the fulfillment of the indicated condition and the right one describes the case when the condition is not met. 
On the line (III) $J(0)=K(0)$, the boundary of the stability condition (see Eq.~(\ref{eq:PStab})) is achieved or the densities of thermodynamic potential (see Eqs.~(\ref{eq:PPot}) and (\ref{eq:FPot})) are equal for (P) and (F) phases. On the line (IV), the densities of the thermodynamic potential (see Eqs.~(\ref{eq:FPot}) and (\ref{eq:QPot})) are equal.
It is worth noting that according to the employed Bogoliubov approximation, the density of condensed atoms $n_{i}$ in any phase must be close to the total density $n$, $n-n_{0i}\ll n$. 

All phases studied above are characterized by different magnetization (see 
Eqs.~(\ref{eq:FMagn}), (\ref{eq:QMagn}), and (\ref{eq:PMagn})), which is 
experimentally determined by measuring the spin populations \cite{Dalibard2012}. 
If the quadrupole degrees of freedom are omitted ($K(0)=0$), then, according to 
Figs.~\ref{fig:ph_diagram_1}, \ref{fig:ph_diagram_2}, the phases are determined 
on the vertical axis $J(0)/|U(0)|$ by the interaction parameter $J(0)$, as 
usually assumed. However, if the parameter $K(0)$ associated with quadrupole 
degrees of freedom is significant, then for a given $J(0)$ the observed phase may 
occur to be different. For example, as one can see from the left panel of 
Fig.~\ref{fig:ph_diagram_2}, we expect to discover a paramagnetic state if the 
interaction is specified, as usually, by $J(0)$ only, but we observe a 
ferromagnetic state due to the presence of quadrupole degrees of freedom. 
Note also that it is possible to study the components of the quadrupole 
matrix, which is specified by the quadrupole operators, by squeezing the spin and 
nematic (quadrupole) variables \cite{Hamley2012}.





\section{Summary}
We have obtained and analyzed a pairwise interaction Hamiltonian for a many-body system of spin-1 atoms in the context of studying interaction effects in ultracold gases. The resulting Hamiltonian includes eight Gell-Mann generators of the SU(3) group: three generators are associated with three components of a spin-1 operator, while the remaining five represent the quadrupole operators, which specify the quadrupole matrix describing the anisotropy of spin fluctuations. We have shown that the quadrupole degrees of freedom are irrelevant for ultracold gases with local interatomic interaction parameterized by the scattering length. However, they should be taken into account if the interaction is considered to be of non-local type. Next, we have applied the obtained interaction Hamiltonian to study the ground-state structure and corresponding single-particle excitations of a weakly interacting gas of spin-1 atoms with Bose-Einstein condensate in a magnetic field. This system exhibits three different types of magnetic ordering: ferromagnetic, quadrupolar, and paramagnetic. The basic thermodynamic quantities such as the ground state thermodynamic potential, pressure, single-particle excitations, and speed of sound are determined and analyzed for each phase. The phase diagram of the system is analyzed and the role of non-local interaction in ultracold gases is discussed.

\section*{Acknowledgements}
The authors are grateful to Andrii Sotnikov for fruitful discussions. The authors acknowledge funding by the National Research Foundation of Ukraine, Grant No.~0120U104963 and the Ministry of Education and Science of Ukraine, Research Grant No.~0120U102252. 

\appendix
\section{Gell-Mann matrices and SU(3) algebra}
\setcounter{section}{1}
The Gell-Mann matrices representing the generators of the SU(3) group are given by
\begin{eqnarray}
\lambda^{1}=\left(
  \begin{array}{ccc}
    0 & 1 & 0 \\ 
    1 & 0 & 0 \\ 
    0 & 0 & 0 \\
  \end{array}
\right), \quad
\lambda^{2}=\left(
  \begin{array}{ccc}
    0 & -i & 0 \\
    i & 0 & 0 \\
    0 & 0 & 0 \\
  \end{array}
\right), \quad
\lambda^{3}=\left(
  \begin{array}{ccc}
    1 & 0 & 0 \\ 
    0 & -1 & 0 \\ 
    0 & 0 & 0 \\
  \end{array}
\right), \nonumber \\
\lambda^{4}=\left(
  \begin{array}{ccc}
    0 & 0 & 1 \\
    0 & 0 & 0 \\
    1 & 0 & 0 \\
  \end{array}
\right), \quad
\lambda^{5}=\left(
  \begin{array}{ccc}
    0 & 0 & -i \\ 
    0 & 0 & 0 \\ 
    i & 0 & 0 \\
  \end{array}
\right), \quad
\lambda^{6}=\left(
  \begin{array}{ccc}
    0 & 0 & 0 \\
    0 & 0 & 1 \\
    0 & 1 & 0 \\
  \end{array}
\right), \nonumber \\
\lambda^{7}=\left(
  \begin{array}{ccc}
    0 & 0 & 0 \\
    0 & 0 & -i \\ 
    0 & i & 0 \\
  \end{array}
\right), \quad
\lambda^{8}=\frac{1}{\sqrt{3}}\left(
  \begin{array}{ccc}
    1 & 0 & 0 \\
    0 & 1 & 0 \\
    0 & 0 & -2 \\
  \end{array}
\right).\label{eq:AGellMann}
\end{eqnarray}
The above matrices, being Hermitian and traceless, have the following property:
\begin{equation}\label{eq:A2}
{\rm Tr}\lambda^{a}\lambda^{b}=2\delta_{ab}
\end{equation}
and meet the following permutation relations:
\begin{equation}\label{eq:A3}
[\lambda^{a},\lambda^{b}]=2if^{abc}\lambda^{c},
\end{equation}
where $f^{abc}$ are the structure constants of the SU(3) group. From Eq.~(\ref{eq:A2}), one obtains
$$
f^{abc}=-\frac{i}4{\rm Tr}\,\lambda^{c}[\lambda^{a},\lambda^{b}],
$$
whence
\begin{equation} \label{eq:A4}
f^{abc}=-f^{bac}=f^{bca}.
\end{equation}
The structure constants $f^{abc}$ have the following numerical values:
\begin{eqnarray}
\fl
\label{eq:A5}
f^{123}=1, \quad f^{147}=-f^{156}=f^{246}=f^{257}=f^{345}=-f^{367}=\frac12, \quad f^{456}=f^{678}=\frac{\sqrt{3}}{2}.
\end{eqnarray}
All other numerical values of $f^{abc}$ not related to the indicated above by permutation are
zero. The anticommutator of the Gell-Mann matrices, as well as the commutator, is linear in $\lambda_{a}$:
\begin{equation} \label{eq:AAntiCom}
\{\lambda^{a},\lambda^{b}\}=\frac43\delta_{ab}+2d^{abc}\lambda^{c},
\end{equation}
where the coefficients $d^{abc}$, symmetric over all indices, are given by
$$
d^{abc}=\frac14{\rm Tr}\,\lambda^{c}\{\lambda^{a},\lambda^{b}\}.
$$
The following values of $d^{abc}$ are different from zero:
\begin{eqnarray}
d^{118}=d^{228}=d^{338}=-d^{888}=\frac{1}{\sqrt{3}}, \nonumber \\
d^{146}=d^{157}=d^{256}=d^{344}=d^{355}=-d^{247}=-d^{366}=-d^{377}=\frac12, \nonumber \\
d^{448}=d^{558}=d^{668}=d^{778}=-\frac{1}{2\sqrt{3}}. \label{eq:A8}
\end{eqnarray}

\section{Solving equation for the vector order parameter}

We recast Eq.~\eqref{eq:MinCond} as the following coupled equations:
\begin{numparts}
\label{eq:state}
\begin{eqnarray}
	\label{eq:state:a}
	a
	\zeta_x
	+
	c
	\left(
	\zeta_x^2
	+
	\zeta_y^2
	+
	\zeta_z^2
	\right)
	\zeta_x^*
	-
	i
	h
	\zeta_y
	=
	0,
	\\\label{eq:state:b}
	a
	\zeta_y
	+
	c
	\left(
	\zeta_x^2
	+
	\zeta_y^2
	+
	\zeta_z^2
	\right)
	\zeta_y^*
	+
	i
	h
	\zeta_x
	=
	0,
	\\\label{eq:state:c}
	a
	\zeta_z
	+
	c
	\left(
	\zeta_x^2
	+
	\zeta_y^2
	+
	\zeta_z^2
	\right)
	\zeta_z^*
	=
	0,
\end{eqnarray}
\end{numparts}
where 
$
	a=		
	\mu
	-
	n_0
	U(0)
	-
	\frac13
	n_0
	K(0)
	-
	n_0
	J(0)
$, 
$	
	c=
	n_0
	\left(
		J(0)
		-
		K(0)
	\right)
$.
The next steps are the following:  
\begin{itemize}
\item to replace \eqref{eq:state:a} by the sum of \eqref{eq:state:a} multiplied on $\frac1{\sqrt{2}}$ and \eqref{eq:state:b} multiplied on $\frac i{\sqrt{2}}$;
\item to replace \eqref{eq:state:b} by the difference of \eqref{eq:state:a} multiplied on $\frac1{\sqrt{2}}$ and \eqref{eq:state:b} multiplied on $\frac i{\sqrt{2}}$.
\end{itemize}
Thus, we obtain:
\begin{numparts}\label{eq:state_mp}
\begin{eqnarray}
\label{eq:state_mp:a}
	(a-h)
	\zeta_{-}
	+
	c
	\left(
	2
	\zeta_{-}
	\zeta_{+}
	+
	\zeta_z^2
	\right)
	\zeta_{+}^*
	&&=
	0,
	\\
\label{eq:state_mp:b}
	(a+h)
	\zeta_{+}
	+
	c
	\left(
	2
	\zeta_{-}
	\zeta_{+}
	+
	\zeta_z^2
	\right)
	\zeta_{-}^*
	&&=
	0,
	\\	
	\label{eq:state_mp:c}
	a
	\zeta_z
	+
	c
	\left(
	2
	\zeta_{-}
	\zeta_{+}
	+
	\zeta_z^2
	\right)
	\zeta_z^*
	&&=
	0,
\end{eqnarray}
\end{numparts}
where
$$
\zeta_x=\frac1{\sqrt{2}}(\zeta_{+}+\zeta_{-}), \quad \zeta_y=\frac i{\sqrt{2}}(\zeta_{+}-\zeta_{-}).
$$

If $\zeta_z=0$, then Eq.~\eqref{eq:state_mp:c} is automatically satisfied and we have
\begin{eqnarray*}
	\left(
		a
		-
		h
		+
		2c
		\left|
		\zeta_{+}
		\right|
		^2
	\right)
	\zeta_{-}
	=0,
\\
	\left(
		a
		+
		h
		+
		2c
		\left|
		\zeta_{-}
		\right|
		^2
	\right)
	\zeta_{+}
	=0.
\end{eqnarray*}
The solutions read 
\begin{eqnarray*}
    a=\pm h,
\qquad
    &&
    \zeta_z=\zeta_{\pm}=0,\quad\zeta_{\mp}=1
,\\
    a=-c,
\qquad
    &&
	\zeta_z=0
	,\quad
	\left|
		\zeta_{\pm}
	\right|
	^2
	=
	\frac12
	\left(
		1
		\pm
		\frac{h}{c}
	\right).
\end{eqnarray*}
Note that the solutions in both lines depend on the direction of a magnetic field along $z$-axis. The solutions in the first and second lines correspond to ferromagnetic and paramagnetic phases (see Eqs.~(\ref{eq:FChem}), (\ref{eq:PChem})), respectively.

In case when $\zeta_{\pm}=0$ but $\zeta_z\neq 0$, we automatically get $\zeta_{\mp}=0$, respectively, and $\zeta_z=1$ ($a=-c$). This solution describes the quadrupolar phase (see Eq.~(\ref{eq:QChem})). The most nontrivial case, when all $\zeta_i\neq 0$ with $i=\{+,-,z\}$, is possible only if $h=0$. This becomes evident if we express $\zeta_z^2$ from the sum of Eq.~\eqref{eq:state_mp:a} multiplied on $\zeta_{+}$ and Eq.~\eqref{eq:state_mp:b} multiplied on $\zeta_{-}$ with subsequent solving Eq.~\eqref{eq:state_mp}. Therefore, at nonzero magnetic field, all solutions of Eqs.~\eqref{eq:state} are found.

\section*{References}
\bibliographystyle{iopart-num}
\bibliography{bib}
\end{document}